\documentclass[%
 reprint,
 amsmath,amssymb,
 aps, nofootinbib,
]{revtex4-1}

\usepackage{graphicx}% Include figure files
\usepackage{dcolumn}% Align table columns on decimal point
\usepackage{bm}% bold math

\usepackage{amsmath,amssymb,amsthm, graphicx, mathrsfs, pgf,tikz,marvosym,fontawesome,slashed, mathtools,array,bbm,yfonts, adjustbox, adforn, fourier-orns, relsize,stmaryrd,simplewick,youngtab,ytableau,youngtab, epigraph, url, bclogo,arydshln,multirow,dsfont, lipsum, tensor,feynmf, wrapfig, natbib, enumitem,subfig}
\usepackage{physics}
\usepackage[left=20mm,right=20mm,top=35mm,bottom=35mm,columnsep=15pt]{geometry} 
\usepackage{placeins}
\usepackage[T1]{fontenc}
\usepackage[utf8]{inputenc}
\usepackage{csquotes}
\usepackage[nottoc]{tocbibind}
\usetikzlibrary{arrows}
\usetikzlibrary{positioning}
\usepackage[english]{babel}

\usepackage[tiny,center]{titlesec}
\titleformat{\section}{\bf}{\thesection}{1em}{\MakeUppercase}  
\usepackage{cancel}
\usepackage[normalem]{ulem}

\definecolor{r}{cmyk}{1,.50,0,.20} 
\usepackage[colorlinks=true, linkcolor=r, urlcolor=r, linktocpage=true, citecolor=r]{hyperref}

\usepackage[titletoc,title]{appendix}

\numberwithin{equation}{section}

\newcommand{\bea}{\begin{eqnarray}}
\newcommand{\eea}{\end{eqnarray}}
\newcommand{\be}{\begin{equation}}
\newcommand{\ee}{\end{equation}}
\newcommand{\dif}{\mathrm{d}}
\newcommand{\osun}{\odot}
\def\nn{\nonumber}

\newcommand{\cR}{\mathcal{R}}

\newcommand{\cA}{\mathcal{A}}

\newcommand{\cT}{\mathcal{T}}
\newcommand{\cK}{\mathcal{K}}
\newcommand{\cO}{\mathcal{O}}

\setcitestyle{square,comma}

\begin{document}

\title{\textbf{Modeling and detecting resonant tides of exotic compact objects}}

\author{Kwinten Fransen}
\email{kwinten.fransen@kuleuven.be}
\affiliation{Instituut  voor  Theoretische  Fysica,  KU Leuven, Celestijnenlaan  200D, B-3001  Leuven, Belgium.}

\author{Gideon Koekoek}
\email{gideon.koekoek@maastrichtuniversity.nl}
\affiliation{Maastricht University, P.O. Box 616, 6200 MD Maastricht, The Netherlands.}
\affiliation{Nikhef, Science Park 105, 1098 XG Amsterdam, The Netherlands.}

\author{Rob Tielemans}
\email{rob.tielemans@kuleuven.be}
\affiliation{Instituut  voor  Theoretische  Fysica,  KU Leuven, Celestijnenlaan  200D, B-3001  Leuven, Belgium.}

\author{Bert Vercnocke}
\email{bert.vercnocke@kuleuven.be}
\affiliation{Instituut  voor  Theoretische  Fysica,  KU Leuven, Celestijnenlaan  200D, B-3001  Leuven, Belgium.}

\begin{abstract}
The event horizon of a black hole in general relativity absorbs all infalling radiation. Any observation of the contrary would immediately challenge the expectation that astrophysical black holes are described by the vacuum Kerr geometry. If a putative black hole does reflect part of the ingoing radiation, its quasinormal mode structure is drastically altered. Low frequency modes can be introduced that are resonantly excited during the inspiral of a binary system. We study the resulting phase shift of the gravitational wave signal. Building on neutron star results, we obtain a model-independent expression for the phase shift that depends only on quasinormal modes and Love numbers of the compact object. We find that the phase shift might be detectable with Einstein Telescope for asymmetric binaries in high signal-to-noise events ($\sim 10^3$), but by far cannot explore the Planck scale. 
\end{abstract}

\maketitle

\thispagestyle{empty}

\onecolumngrid
\tableofcontents
\vspace{1cm}
\twocolumngrid

\newpage

\section{Introduction and summary} \label{sec: Introduction and summary} 
A fundamental question of present-day research concerns the nature of astrophysical black holes: are they truly described by the Kerr family of General Relativity? The ever-increasing precision of gravitational wave experiments is a sensitive probe that offers insight into this question. A rich phenomenology has sprouted in recent years on horizonless compact objects, which have some structure close to the would-be event horizon, and can serve as testable alternatives to the Kerr paradigm. Many of the models for such exotic compact objects (ECO) are directly inspired or even predicted by fundamental physics models that describe new physics near black holes, ranging from dark matter to quantum gravity \cite{Morris1988, Mazur2001, Schunck:2003kk, Mathur2005, Barcelo2007, Barcelo:2010vc, Barcelo:2014cla, Holdom:2016nek, Raposo:2018xkf, Liebling:2012fv, Cardoso:2019rvt, PhysRevLett.66.1659, Gleiser:1993pt, Seidel:1993zk}. Exploring their gravitational wave signatures transforms compact merger events into fundamental physics laboratories.

A main theme in this exploration is to find model-independent gravitational wave (GW)  observables that can distinguish an exotic compact object from a black hole. Two key sets of observable that have been studied recently are GW echoes following the merger and tidal effects during inspiral.  GW echoes \cite{Cardoso2016(2),Cardoso2016,Cardoso2017,Conklin:2017lwb,Mark:2017dnq,Price:2017cjr,Wang:2018mlp,Wang:2019szm,Bueno:2017hyj,Ashton:2016xff,Maggio:2019zyv} arise when the ECO reflects part of the incoming gravitational radiation: the interaction between the ECO and its light ring behaves as a resonant cavity inducing a particular set of trapped quasi-normal modes (QNM) of significantly lower frequency than the standard black hole QNMs. Potentially observable with LIGO/Virgo, the data analysis hunt is on, with no conclusive evidence up to date \cite{Abedi:2016hgu,Abedi:2017isz,Ashton:2016xff,Uchikata:2019frs,Lo:2018sep,Nielsen:2018lkf,Testa:2018bzd,Tsang:2018uie,Westerweck:2017hus}.  Adiabatic tidal interactions on the other hand can lead to deformability and heating effects which typically appear in the GW signal  as highly suppressed corrections during most of the inspiral. Those effects are in principle detectable with 3G detectors \cite{Cardoso:2017cfl,Maselli:2017cmm,Johnson-McDaniel:2018uvs,Maselli:2018fay,Datta:2019epe,Datta:2020gem,Pani2010,Cardoso:2019nis,Pons2002,Gualtieri2001,Flanagan2007,Cardoso2017(2),Binnington2009,flanagan2008,Damour2009,Cardoso:2019nis,Datta:2019euh,Ruoff2000,Kokkotas1995}. 

The same physical mechanism underlying GW echoes offers the possibility for another effect in the inspiral phase: low frequency QNMs can be excited by the driving of the companion. At resonance, tidal effects suddenly become more dominant. This gives two potentially observable effects. First, sharp peaks in the emitted gravitational wave power are to be expected when the orbital frequency matches the resonant frequency of an internal oscillation mode of the object. This effectively leads to a high frequency `glitch' in the observed signal which is most likely inaccessible to current or planned detectors \cite{Cardoso:2019nis}. Second, the significant enhancement of energy in an internal mode causes a phase difference in the GW signal. This kind of resonance phenomenon has already been discussed at length in the context of neutron stars (NS) \cite{Kojima1987, Reisenegger1993, Lai1994, Pons2002} and white dwarfs \cite{Rathore:2004gs}, and numerically for particular ECO models such as gravastars and boson stars \cite{Cardoso:2017cfl,Cardoso2017,Pani2010, Macedo2013, Cardoso2016}. A recent model-independent data analysis for the phase shift due to inspiral resonances showed no deviation from GR in second-generation GW observations yet \cite{NikhefResonances}. However, a detailed treatment for the form and expected size of the phase shift for generic compact objects is currently not available.

In this paper, we investigate specifically the perspectives for detection of the phase shift induced by a resonant excitation of the heaviest object in a comparable mass binary merger. The companion could be a black hole, a neutron star or even another ECO although tidal effects in the companion will not be taken into account.
By extrapolating known results for neutron stars, we write down the phase shift in terms of QNMs and Love numbers of the ECO.
We  determine the prospects for detection of those modes using ground-based GW detectors using a Fisher analysis. Although our setup is more generally valid, we do this analysis for the simplest model of a reflecting surface at coordinate radius $r_0 = 2M_1 (1 + \epsilon)$ where $M_1$ is the ECO mass, $M_2$ will be the companion mass, and  $\epsilon$ is a dimensionless parameter controlling the compactness of the ECO or alternatively a ``closeness'' parameter that indicates how close the reflective surface is to the would-be horizon \cite{Cardoso:2019rvt}. In particular, $\epsilon$ is chosen such that the surface lies well within the photonsphere but is, in proper distance, more than a Planck length away from the putative horizon. For a hundred solar mass object, this implies $\epsilon$ lies roughly between $10^{-2}$ and $10^{-80}$.
We find that with current detector capacity observation of the phase shift of our model is ruled out. With third generation GW observatories, things look better but might still be out of reach. We focus on Einstein telescope and find that only at high signal-to-noise ratios of $\sim 10^3$ the phase shift becomes detectable for mass ratios of roughly $M_2/M_1 \lesssim 10^{-2}$ and a wide range of $\epsilon$, but nevertheless corresponding to a proper distance away from the horizon of many orders of magnitude above Planck scale. The phase shift at leading order in $\epsilon$ scales with the inverse of the mass ratio, suggesting better prospects for extreme mass ratio inspirals (EMRI) with LISA. However, our approximations do not extend to that setup and we defer a proper EMRI study to future work.
Along the way, we also discuss how the resonance can be written in an effective theory as is done for GW echoes. We explore the post-Newtonian structure of this approach and, thereby, clarify how the dynamic tidal deformability induces a difference in gravitational wave emission between an ECO and a black hole at resonance. 

The rest of the paper discusses our main points as follows. In section \ref{sec: HO}, we discuss the inspiral of a featureless point particle into a compact object. After a quick review of the linear response to the tidal field of the companion, we model the ECO as a point particle dressed with multipolar deformation degrees of freedom.  Such a description is applicable irrespective of the details of the object and has been developed and applied previously for stellar objects \cite{Goldberger2004}. Following this previous literature  we derive the phase shift as consequence of a resonance in the GW signal by estimating the orbital energy that leaks into a specific mode of oscillation.  For neutron stars, this driving is known to be related to the  overlap integrals that describe the internal structure of the object \cite{Flanagan2007, Chakrabarti2013u1306}. By assuming only the fundamental mode contributes, we express the overlap integrals in terms of the Love numbers to arrive at a largely model-independent estimate of the phase shift, see equation \eqref{eqn:PS} below.

In section \ref{sec:nearzone}, we discuss the essential difference with stellar objects in the form of the tidal response function. This requires input depending on the nature of the object itself. We give a near-zone analysis of the object, closely following the discussion in \cite{Mark:2017dnq} developed for GW echoes, and characterize an ECO by its boundary conditions that replace the purely absorbing black hole horizon. We show how, in a low frequency limit, the transfer function ${\cK}(\omega)$ of \cite{Mark:2017dnq} is proportional to the linear response function $F(\omega)$ of the effective theory of quadrupole deformations, elucidating the relation to the high-frequency glitch described in \cite{Cardoso:2019nis} and tidal effects in the post-Newtonian expansion. For calculational reasons, we focus on the EMRI limit in this section.

In section \ref{sec: dectectability}, we first discuss the basic conditions that have to be satisfied for a resonance to be seen in a gravitational wave signal, including relation to mass and details of central object and ECO. We restrict to a simple model and qualitatively observe that such a detection is possible in principle. However, we then perform a Fisher analysis and find that the resulting phase shift is unlikely to be seen by the Einstein Telescope. 

We conclude in section \ref{sec: outlook} that, even though our results are only a first order of magnitude estimate, they serve as an indication that a more detailed analysis of the extreme-mass ratio limit is worthwhile.
Appendix  \ref{app:effectivetheory} and \ref{app:near} contain technical details relevant to sections \ref{sec: HO} and \ref{sec:nearzone} respectively.

\section{Harmonic oscillator model of a compact object} \label{sec: HO}

We first give a quick recap of the phase shift derived using the Newtonian approximation \cite{Lai1994, Reisenegger1993, Flanagan2007} and the harmonic oscillator model used to describe the linear response of stellar objects subject to a tidal field \cite{Chakrabarti2013,Chakrabarti2013u1306,steinhoff2016}. We end by expressing the phase shift in terms of the GR Love numbers and mode frequencies.

The reader interested in the results can jump to section \ref{subsection:ECOphaseshift}.

\subsection{Phase shift in the Newtonian approximation}

We consider two masses $M_1$ and $M_2$ with mass ratio $q = M_2/M_1$.  In the early inspiral, the motion is dominated by the point-particle motion at the leading (post-)Newtonion order. On the additional assumption of quasi-circular orbits, it is described by the relative separation of the masses, $r(t)$, and the orbital phase, $\phi(t)$:
\begin{subequations}
	\begin{align}
	&r(t) = \frac{4}{5^{3/2}}\frac{(5\mathcal{M}_{\rm c})^{5/4}}{\mu^{1/2}}|t_{\rm c}-t|^{1/4}, \\
	&\phi(t) = \phi_{\rm c}-\left(\frac{t_{\rm c}-t}{5\mathcal{M}_{\rm c}}\right)^{5/8}, \label{eq: orbital phase} 
	\end{align}\label{eq:randphi}%
\end{subequations}%
in which $t_{\rm c}$ and $\phi_{\rm c}$ are the time and phase of coalescence respectively,  $\mathcal{M}_{\rm c}$ is the chirp mass and $\mu$ the reduced mass given by
\begin{align}
\mathcal{M}_{\rm c} = \frac{(M_1M_2)^{\frac{3}{5}}}{(M_1+M_2)^{\frac{1}{5}}}, && \mu = \frac{M_1M_2}{M_1+M_2}.
\end{align}

Now consider $M_1$ to be an ECO that can be equipped with additional internal degrees of freedom, $M_2$ will still be modeled as a featureless point particle. A resonant excitation of an internal degree of freedom during the inspiral effectively causes a phase shift in the gravitational wave signal; the GW phase $\Phi(t)$ away from the resonance regime is given by \cite{Flanagan2007}
\begin{equation}
\Phi(t)=\begin{cases} \Phi_{\rm pp}(t) & \text{if } t-t_0\ll -\Delta t \\ \Phi_{\rm pp}(t) + \left(\dfrac{\dot{\phi}(t)}{\dot{\phi}_{\rm R}}-1\right)\Delta\Phi
& \text{if } t-t_0\gg \Delta t, \end{cases}
\label{FlanaganEquation}
\end{equation}
where $\Delta t$ estimates the duration of resonance, $\dot{\phi}_{\rm R}$ is the angular velocity of the binary evaluated at resonance $t=t_0$ and $\Phi_{\rm pp}(t)$ is the GW phase as predicted for a point-particle. Our order-of-magnitude estimate for the phase shift $\Delta\Phi$ is
\begin{equation}
\Delta\Phi \approx \left.2\dot{\phi}\frac{\Delta E_{n\ell m}}{\dot{E}_{\rm GW}}\right|_{t=t_0}, \label{eq: estimate phase shift}
\end{equation}
in which $\dot{E}_{\rm GW}= 32/5\:(\mathcal{M}_{\rm c}\dot{\phi})^{10/3}$ is the `Newtonian' gravitational wave luminosity and $\Delta E_{n\ell m}$ denotes the orbital energy loss in a particular, resonantly excited, $(n,\ell,m)$-multipole mode. The second factor is the time scale associated to the energy loss $\Delta E_{n \ell m}$ during normal GW emission hence $\Delta\Phi$ estimates the shift in GW phase when passing through a resonance epoch. We will obtain the energy loss $\Delta E_{n \ell m}$ from a linearized harmonic oscillation model put forward in  \cite{Chakrabarti2013u1306} on assuming the no back-reaction approximation (the orbital motion of the companion is a constant supply of energy without the mode oscillations influencing the orbit).

\subsection{Oscillating stars}\label{subsec:harmonic_oscilator_Stars}

Chakrabarti et al. \cite{Chakrabarti2013u1306} have shown how to incorporate multipolar degrees of freedom on the worldline of a point particle to describe a generically deformed object in the Newtonian regime (this will naturally restrict us to tidal effects of electric type). Fundamental quantities are the deformation amplitudes $c_{n\ell m}(t)$ of a specific normal mode of oscillation and the (quasi) normal modes of the central object $\omega_{n\ell}$. Note that the QNMs do not carry an $m$-index because of assumed spherical symmetry of the ECO. In the regime of linear response, these amplitudes are described by a driven, damped harmonic oscillator 
\begin{equation}
\ddot{c}_{n\ell m}+2\gamma_{n\ell}\dot{c}_{n\ell m}+\omega_{n\ell}^2c_{n\ell m} = f_{n\ell m}, \label{eq: HO}
\end{equation}%
where a dot denotes a time derivative and with $\omega_{n \ell}$ the frequency of oscillation $\gamma_{n\ell}$ and damping coefficients; they are related to the quasi-normal frequencies  $\omega^{\text{QNM}}_{n\ell}$ of the central object as
\begin{subequations}
	\begin{align}
	&\gamma_{n\ell} \equiv -\mathrm{Im}\: \omega_{n\ell}^{\text{QNM}}, \\
	&\omega_{n\ell }^2\equiv\left(\mathrm{Re}\:\omega_{n\ell}^{\text{QNM}}\right)^2+\left(\mathrm{Im}\: \omega_{n\ell}^{\text{QNM}}\right)^2. \label{eq2: omega}
	\end{align}
\end{subequations}%

The driving term $f_{n\ell m}$ describes how the internal degrees of freedom of the central object couple to the external tidal field of the companion \cite{Chakrabarti2013u1306}. The precise relation is most easily written in a symmetric trace-free (STF) tensor basis $f_{n\ell m} \to \hat{f}_{n K_{\ell}}$ where $K_\ell$ is a multi-index $K_\ell = k_1k_2\ldots k_\ell$ and the hat denotes STF projection, see section IV and/or appendix A of \cite{Chakrabarti2013u1306} for a more detailed description. The driving term is given by 
\begin{equation}
\hat{f}_{nK_\ell} = -\frac{I_{n\ell}}{\ell!}\hat{\partial}_{K_\ell} \Phi(\mathbf{r}). \label{eq: driving}
\end{equation}
In this expression, $I_{n\ell}$ is the overlap integral encoding equation of state information of the central object, the hatted $\partial_{K_{\ell}}$ refers to the STF projection of a multi-index partial derivative\footnote{For instance, applying this notation to a partial differential $\hat{\partial}_{K_2}= \frac{1}{2}(\frac{\partial}{\partial r^{k_1}}\frac{\partial}{\partial r^{k_2}}+\frac{\partial}{\partial r^{k_2}}\frac{\partial}{\partial r^{k_1}})-\frac{\delta_{k_1 k_2}}{3} \frac{\partial}{\partial r^{i}}\frac{\partial}{\partial r_{i}}$, with $k_i$ taking values in $1$,$2$, $3$ and $r^{k_1}$ the Cartesian coordinates of $\mathbf{r}$. This is, however, simply equal to $\frac{\partial}{\partial r^{k_1}}\frac{\partial}{\partial r^{k_2}}$ when acting on a Newtonian potential in vacuum.}   and $\Phi(\mathbf{r})$ is the Newtonian gravitational potential of the companion evaluated at its position $\mathbf{r}=\mathbf{r}(t)$, in particular
\begin{equation}
\Phi(\mathbf{r}) = -\frac{M_2}{|\mathbf{r}(t)|}. \label{eq: potential}
\end{equation}
We will focus only on quasi-circular orbits in the equatorial plane. This simplifies the driving term  to the split form
\begin{subequations}
	\begin{align}
	f_{n\ell m} = F_{n \ell m}(t)e^{-im\phi(t)}, 
	\label{eqn:f}
	\end{align}%
	where the amplitude is given by
	\begin{align}
	F_{n\ell m}(t) = \mathcal{N}_{\ell m}\frac{M_2|I_{n\ell}|}{[r(t)]^{\ell+1}},
	\end{align}\label{eq: driving cirular orbit}%
	and where $r(t)$ and $\phi(t)$ are given in \eqref{eq:randphi} and $\mathcal{N}_{\ell m}$ is a numerical prefactor
	\begin{align}
	\begin{aligned}
	\mathcal{N}_{\ell m}= &\frac{(-1)^\ell 2^{m-1}}{\Gamma\left(\frac{-\ell-m+1}{2}\right)\Gamma\left(\frac{\ell-m}{2}+1\right)}\\&\quad\times\sqrt{\frac{8\pi(2\ell-1)!!}{2\ell!}\frac{\Gamma(\ell-m+1)}{\Gamma(\ell+m+1)}}.%
	\end{aligned} \label{eq: Nlm}%
	\end{align}%
\end{subequations}%
Note that we take the modulus of the overlap integral when compared to \eqref{eq: driving}. Any phase corresponding to $I_{n\ell}$ can be absorbed in the definition of the orbital phase \eqref{eq: orbital phase}.

When a solution to the oscillator equation \eqref{eq: HO} is obtained, the energy stored in the $(n,\ell,m)$-mode can be obtained in the standard fashion from the solution of \eqref{eq: HO} as follows
\begin{equation}
E_{n\ell m}(t) = \frac{1}{2}\left(\mathrm{Re}\:\dot{c}_{n\ell m}(t)\right)^2+\frac{\omega_{n\ell}^2}{2}\left(\mathrm{Re}\:c_{n\ell m}(t)\right)^2.
\end{equation}%
The energy in the modes after passing through a resonance can be approximated by \cite{Rathore:2004gs,Reisenegger1993}
\begin{align}
E_{n\ell m}(t) \approx \Delta E_{n\ell m}e^{-2\gamma_{n\ell}(t-t_0)}\theta\left(t-(t_0+\Delta t)\right),
\end{align}
where $\theta$ is the Heaviside step function given by
\begin{equation}
\theta(t) = \begin{cases}1&\text{if } t\geqslant 0\\ 0&\text{if } t<0 \end{cases},
\end{equation}
and
\begin{equation}
\Delta E_{n\ell m}= \frac{\pi F_{n\ell m,\rm R}^2}{4|m|\ddot{\phi}_{\rm R}}. \label{eq: absorption energy}
\end{equation}
Here, $F_{n\ell m,\rm R}$ and $\ddot{\phi}_R$ are respectively the driving amplitude, see \eqref{eqn:f}, and change in angular velocity of the binary evaluated at resonance. Note that $\Delta E_{n\ell m}$ gives the \textit{total} amount of energy transferred from the orbit to the mode while $E_{n\ell m}(t)$ is the \textit{present} amount of energy in the mode (that is, the total amount \textit{minus} the amount which is dissipated by internal friction or gravitational wave emission). Using this result \eqref{eq: absorption energy} for the phase shift estimate \eqref{eq: estimate phase shift} yields 
\begin{equation}
\Delta \Phi = \frac{25\pi\mathcal{N}_{\ell m}^2}{6144|m|^{\frac{1}{3}(4\ell-11)}}M_1^{-\frac{2}{3}\ell-\frac{11}{3}}\omega_{n\ell}^{\frac{4}{3}\ell-\frac{14}{3}} |I_{n\ell}|^2. \label{eq:DeltaPhi_stars}
\end{equation}
This expression depends on the overlap integrals and hence on the internal structure and mass distribution of the central object $M_1$.

%wasn't necessary not II14 in the previous is the next due to introdcution \theta ...

\subsection{From stars  to compact objects}\label{subsection:ECOphaseshift}

To make contact to ECOs, we express the overlap integrals in \eqref{eq:DeltaPhi_stars} in terms of the Love numbers using the  effective theory of tidal deformations. To this end, we will restrict to fundamental $n=1$ modes\footnote{We choose the convention $n=1$ for the fundamental mode as it is convenient to describe the trapped QNM's in \eqref{eq: QNM frequency}. However, this means a shift is required in the comparison with the traditional $n=0$ choice for stellar oscillations.} and omit the subscript $n$ henceforth. 

For quadrupolar deformations in GR, one can construct an effective action for a dynamical quadrupole degree of freedom $Q^{ab}$ on the wordline of a point particle (in our case: the ECO). The linear response to an external tidal field is given through the following coupling with the electric component of the Weyl tensor $E^{ab}$, in the frequency domain
\begin{equation}
\tilde Q^{ab}= - \frac 12 \tilde F(\omega) \tilde{E}^{ab}, \label{eq: quadrupole linear response}
\end{equation}
where we have used $\tilde Q^{ab}$, $\tilde{E}^{ab}$  to indicate the Fourier transforms of $Q^{ab}$, $E^{ab}$ and $\tilde{F}(\omega)$ is the linear response function, in the frequency domain, that determines the strength of the tidal interaction and depends on the properties of the compact object, such as its equation of state.

In the low-frequency regime, the Taylor expansion of $\tilde F(\omega)$ gives the tidal constants:
\begin{equation}
\tilde{F}(\omega) = \mu_2 + i \lambda \omega+2\mu_2' \omega^2 +\ldots\,,\label{eq:linearresponse_lovenumbers}
\end{equation}
with $\mu_2 = \frac{2M_1^5}{3} k_2$ for the relativistic quadrupolar dimensionless electric tidal Love number $k_2$, $\lambda$ related to absorption and $\mu_2'$ parametrizing tidal response beyond the adiabatic limit \cite{Chakrabarti2013}. Note that the relativistic tidal Love numbers of non-spinning objects can be split in to two classes based on parity: electric (even-parity) and magnetic (odd-parity). We only consider the electric type as they also exist in the Newtonian theory, in contrast to the magnetic-type Love numbers \cite{Pani:2018inf}. Furthermore, we use the convention for $k_2$ of \cite{Cardoso:2017cfl}, where it is shown explicitly how this convention differs, for example, from the conventions of \cite{Binnington2009,Hinderer:2007mb}. Near resonance, the response function has a pole structure, 
\be
\tilde{F}(\omega) \approx \sum_n \frac{I_{n\ell}^2}{\omega^2_{n\ell}-\omega^2}\,,\label{eq:linearresponse_poles}
\ee
with $\omega_{n\ell}$ the resonant frequencies.  As indicated, the residues are related to the  overlap integrals. By equation \eqref{eq:linearresponse_lovenumbers} and \eqref{eq:linearresponse_poles}, we can then immediately relate Love number and overlap integral.

For generic multipoles, we use the Newtonian limit of \cite{Chakrabarti2013u1306}. In appendix \ref{app:effectivetheory}, we list those results and repeat the above reasoning. The main assumption made is that overlap integrals for fundamental modes dominate above their overtones; hence we truncate to $n=1$. The overlap integral is given as:
\begin{equation}
I_{1\ell}^2 = \frac{2\ell!}{(2\ell-1)!!}\omega_{1\ell}^2 M_1^{2\ell+1}k_\ell, \label{eq: overlap - Love number}
\end{equation}
with $k_{\ell}$ the electric mutipolar Love numbers. The phase shift \eqref{eq:DeltaPhi_stars} becomes
\begin{subequations}
	\begin{equation}
	\Delta \Phi = \mathcal{C}_{\ell m}(M_1\omega_\ell)^{\frac{4}{3}(\ell-2)}\frac{|k_\ell|}{q(1+q)^{\frac{1}{3}(2\ell-1)}}, \label{eq:DeltaPhi_stars2(2)}
	\end{equation} \label{eq:DeltaPhi_stars2}%
	where the numerical prefactor $\mathcal{C}_{\ell m}$ is given by
	\begin{equation}
	\mathcal{C}_{\ell m} = \frac{25\pi }{6144|m|^{\frac{1}{3}(4\ell-11)}}\frac{2\ell!}{(2\ell-1)!!}\mathcal{N}_{\ell m}^2.
	\end{equation} \label{eqn:PS}%
\end{subequations}
The numerical factor $\mathcal{N}_{\ell m}$ is given in equation \eqref{eq: Nlm}.

We have made the  assumption that the same relations as for Newtonian stars holds for ECO's. However, the tidal behaviour of ECOs is counterintuitive compared to normal Newtonian fluid objects, in particular the Love numbers for very compact objects can actually become negative (see eq.\ \eqref{eq.:lovenumberlog} below). Therefore we work with the square of the overlap integral and take the modulus in \eqref{eq:DeltaPhi_stars2(2)} to have the correct physical sign of the phase shift. It would be interesting to perform an independent check of the complex nature of the overlap integrals for ECOs by evaluating the residues in the response function $\tilde{F}(\omega)$ explicitly \eqref{eq:linearresponse_poles}.

\section{Resonances of Exotic Compact Objects}
\label{sec:nearzone}
For simplicity, we keep  the discussion focused on a non-rotating ECO. Section \ref{sec:sub1} reviews relevant elements of previous work, mainly \cite{Mark:2017dnq}, while introducing our notation. It also stresses the natural relation between echoes and low frequency resonances. Subsequently, in section \ref{sec:sub2}, we make a connection between that approach and the effective harmonic oscillator model presented in section \ref{sec: HO}. That relation is elegantly summarized, in the low frequency limit, by \eqref{eqn:Llowf}. Finally, in \ref{sec:enhancedflux}, we exhibit the enhanced gravitational wave flux that was also discussed in \cite{Cardoso:2019nis} as well as indicate how it suggests a resummation of the post-Newtonian expansion with dynamical tides. 
\subsection{Perturbing an ECO}
\label{sec:sub1}
Gravitational perturbations around a non-rotating black hole background are governed by the Regge-Wheeler (RW) equation \cite{regge1957}, 
\be
\frac{\dif^2 X_{\ell m}(r,\omega)}{\dif r_*^2}+(\omega^2-V({r})) X_{\ell m}(r,\omega) =  T_{\ell  m}(r,\omega),
\label{eqn:RW}
\ee
with
\be
V(r) = \frac{r(r-2M_1)}{r^4}\left(\ell(\ell+1)-\frac{6 M_1}{r}\right),
\ee
and the tortoise coordinate $r_* = r + 2M_1 \log{(\frac{r}{2 M_1}-1)}$. We will denote the radial coordinate of the horizon as $r_+ = 2 M_1$. The frequency domain Regge-Wheeler function $X_{\ell m}$ can be straightforwardly connected to linearized metric perturbations $h_{\mu \nu}$ around a Schwarzschild background for the odd-parity sector in Regge-Wheeler gauge \cite{Martel:2005ir}

\begin{equation}
X_{\ell m}(r,\omega) = \frac{r-2M_1}{r^2 \ell(\ell+1)\sqrt{2 \pi}} \int d t d \Omega e^{-i\omega t} \bar{P}_{\ell m}^A h_{r A}\,,
\end{equation}

with $d \Omega$, the area form on the unit two-sphere, the index $A$ running over the angular coordinates $\phi, \theta$ and $\bar{P}_{\ell m}^A$ the complex conjugate of the parity-odd vector spherical harmonics, which can be constructed from the ordinary spherical harmonics $Y_{\ell m}$ by

\begin{equation}
P_{\ell m}^A=-\epsilon^{AB}D_B Y_{\ell m}\,,
\end{equation}

with $D_A$, $\epsilon_{AB}$ respectively the covariant derivative and the Levi-Civita tensor on the unit two-sphere. For the even-parity sector, the relation is more complicated but can nevertheless be found through the Chandrasekhar transformation \cite{chandrasekhar1975}. The source term $T_{\ell  m}(r;\omega)$ can similarly be connected to the stress-energy tensor but consider, for the moment, the homogeneous equation \eqref{eqn:RW} with $T_{\ell  m} = 0$. Two independent solutions are given asymptotically by
\begin{subequations}
	\be
	X_{\ell m}^{\text{in}} \sim    \begin{cases}
		e^{-i \omega r_{*}}, & r \to r_+ \\
		A^{\text{out}}  e^{i\omega r_{*}}+A^{\text{in}} e^{-i\omega r_{*}}, & r \to \infty
	\end{cases}
	\ee%
	\be
	X_{\ell m}^{\text{up}} \sim \begin{cases}
		B^{\text{in}}  e^{-i \omega r_{*}}+B^{\text{out}} e^{i \omega r_{*}}, & r \to r_+ \\
		e^{i\omega r_{*}}, & r \to \infty\,.
	\end{cases}%
	\ee
	\label{eqn:asymptotic}
\end{subequations}%

Here, the asymptotically outgoing and ingoing amplitudes, $A^{\text{out}}$, $A^{\text{in}}$ of $X_{\ell m}^{\text{in}}$ as well as the horizon outgoing and ingoing amplitudes $B^{\text{out}}$, $B^{\text{in}}$ of $X_{\ell m}^{\text{up}}$ also depend on $\omega, \ell, m$ although this is left implicit for notational simplicity. The solutions introduced in \eqref{eqn:asymptotic} have the right boundary conditions for a black hole respectively at the horizon and asymptotically. However, for a compact object distinct from a black hole, \eqref{eqn:RW} is only valid up until a particular $r_0 > 2 M_1$ and one must impose an alternative boundary condition. We will denote the homogeneous solution satisfying this boundary condition  $X_{\ell m}^{\text{reg}}$. To go from the black hole to the exotic compact object then essentially amounts to replacing $X_{\ell m}^{\text{in}}$ with $X_{\ell m}^{\text{reg}}$. Since for generic frequencies $X_{\ell m}^{\text{in}}$ and $X_{\ell m}^{\text{up}}$ are independent, one can generally express $X_{\ell m}^{\text{reg}}$ as a linear combination of both. As by assumption $r_0$ is close to $r_+$, we will follow the notation of \cite{Mark:2017dnq} and characterize this homogeneous solution with the correct boundary conditions at the ECO as
\be
X_{\ell m}^{\text{reg}} \propto e^{-i\omega (r^*-r_0^*)}+\cR(\omega) e^{i \omega (r^*-r^*_0)},
\label{eqn:reflectivityBC}
\ee
with $\cR(\omega)$ an arbitrary function of the frequency and the mode numbers $(\ell, m)$. For instance, the specific reflectivity of $\cR = -1$ was discussed in \cite{Cardoso:2019nis}. Note that, despite the apparent remaining freedom, this is a strong reduction of the possible boundary conditions as in general they could be nonlinear functions of all other modes $X_{\ell m}$. 

Given an understanding of the compact object, one can exactly determine what the boundary condition expressed through \eqref{eqn:reflectivityBC} should be by imposing appropriate regularity conditions inside the object and matching this to our exterior description as in the examples of neutron stars \cite{Gualtieri2001}, gravastars \cite{Pani2010} or boson stars \cite{Macedo2013}. Alternatively, one could encode this information by continuing $r$ beyond $r_0$ but with an alternative potential which effectively captures certain interesting features as in \cite{volkel2017}. We will keep the discussion general on the level of this boundary condition following \cite{Mark:2017dnq}.  

Consider \eqref{eqn:RW} in the presence of a source $T_{\ell m}$. Given such a source term, one can solve \eqref{eqn:RW} with variation of parameters using the homogeneous solutions with the appropriate boundary conditions. For a black hole this gives
\begin{subequations}
	\be
	\begin{aligned}
		X^{\text{BH}}_{\ell m} = \frac{1}{W_{\ell m\omega}}&\Big(X_{\ell m }^{\text{up}} \int^{r^*}_{r^+} \dif r_*' X_{\ell m}^{\text{in}}  T_{\ell  m} \\
		&\quad+ X_{\ell m}^{\text{in}} \int^{\infty}_{r^*} \dif r_*' X_{\ell m}^{\text{up}}  T_{\ell  m} \Big),
	\end{aligned}%
	\ee%
	with the wronskian
	\be
	W^{\text{BH}}_{\ell m\omega} = (X_{\ell m}^{\text{in}} \frac{\dif X_{\ell m}^{\text{up}}}{\dif r_*}-X_{\ell m }^{\text{up}} \frac{\dif X_{\ell m}^{\text{in}}}{\dif r_*}).
	\ee%
\end{subequations}
Instead, for the ECO, using the same procedure in combination with \eqref{eqn:reflectivityBC} one finds
\bea
X_{\ell m} &=& X^{\text{BH}}_{\ell m}+ \frac{\cK}{W^{\text{BH}}}  X_{\ell m}^{\text{up}} \int^{\infty}_{-\infty} \dif r_*' X_{\ell m}^{\text{up}}  T_{\ell  m},
\label{eqn:Rsol}
\eea
with
\be
\cK \equiv \frac{\cT_{\text{BH}} \cR e^{-2i\omega r^*_0}}{1-\cR_{\text{BH}} \cR e^{-2i\omega r^*_0}},
\label{eqn:Kmark}
\ee
and with the black hole reflection and transmission amplitudes
\begin{align}
\cR_{\text{BH}}= \frac{B_{\text{in}}}{B_{\text{out}}}, && \cT_{\text{BH}} = \frac{1}{B_{\text{out}}}.
\end{align}
An observer measuring gravitational waves at $r \to \infty$ would see the black hole as
\be
X^{\text{BH}}_{\ell m}(r \to \infty) \sim Z^{\infty}_{\text{BH}}e^{i\omega r_{*}}, 
\ee
with
\be
Z^{\infty}_{\text{BH}} =\frac{1}{W^{\text{BH}}}\int^{\infty}_{-\infty} \dif r_*' X_{\ell m}^{\text{in}}  T_{\ell  m}.
\label{eqn:Zinfty}
\ee
On the other hand, for the exotic compact object we have
\be
X_{\ell m}(r \to \infty) \sim (Z^{\infty}_{\text{BH}}+\cK Z^{H}_{\text{BH}})e^{i\omega r_{*}},
\ee
with
\be 
Z^{H}_{\text{BH}} =\frac{1}{W^{\text{BH}}}\int^{\infty}_{-\infty} \dif r_*' X_{\ell m}^{\text{up}}  T_{\ell  m}.
\label{eqn:ZH}
\ee

Two crucial remarks can be made using the structure of \eqref{eqn:Kmark} \cite{Mark:2017dnq}. The first is that from the expansion 
\be
\cK = \cT_{\text{BH}}\cR e^{-2i\omega r_0^*} \sum_{n=0}^{\infty} (\cR_\text{BH} \cR)^{n-1}e^{-2i(n-1)\omega r_0^*},
\ee
an observer sees an initial perturbation of the ECO decaying as a black hole ringdown followed by a series of echoes spaced by a time interval $2 |r_0^*|$. Secondly, there are a set of QNMs associated to this characteristic timescales arising from the poles in $\cK$, particularly from
\be
0=1-\cR_{\text{BH}} \cR e^{-2i\omega r^*_0}\,.
\label{eqn:QNMcondition}
\ee

Therefore, one easily relates a specific set of resonant modes in ECO and the phenomenology of echoes. Echoes only truly emerge if $r_0$ is sufficiently close to the horizon or more precisely $r_0^* \ll 3M_1$. This implies that the characteristic frequency $\frac{1}{2 |r_0^*|}$ is relatively low and this is for our purpose the crucial difference with black hole QNMs. Indeed, BH QNMs are in principle equally susceptible to resonant excitation but this is of less interest in a generic inspiral because the frequencies are too high to be resonantly excited in that stage of the binary evolution (see however the observation of \cite{Loutrel2017,Nasipak2019} for highly eccentric inspirals). For the ECO discussed here, on the other hand, the QNM frequencies can be significantly lower such that the resonant frequencies can occur during the inspiral.  \\

\subsection{Transfer to tidal response}
\label{sec:sub2}
The proper way to include the companion mass $M_2$ to the previous description, would be to consider it as a near-zone description of $M_1$ and to impose boundary conditions found by a matched asymptotic expansion with an outer-zone\footnote{Sometimes the outer-zone is called the near-zone (also orbital-zone) while our near-zone is called the inner-zone (also body-zone).} containing $M_2$. This outer-zone in turn could for instance be described by a post-Newtonian expansion \cite{mundim2014}. Such a procedure carried out in full, however, becomes cumbersome very quickly and it is more convenient to instead use the near-zone to match to an `intermediary' effective action \cite{steinhoff2016}. The near-zone features of the ECO in this approach are encoded in the effective (linearized) dynamical tidal response $\tilde{F}(\omega)$ as in \eqref{eq: quadrupole linear response}. This tidal response  was related explicitly to the asymptotic form of $X_{\ell m \omega}^{\text{reg}}$ in a $2M_1\omega \ll 1$ expansion by  \cite{Chakrabarti2013}. These authors used the black hole perturbation equations with an effective source to describe the near-zone perturbation equations of neutron stars, but the approach is applicable to more general compact objects. The effective source is determined based on the intermediary effective action by matching the two descriptions in the asymptotic region of the near-zone. This gives the desired relation between $X_{\ell m \omega}^{\text{reg}}$ and $\tilde{F}(\omega)$. Concretely, consider the particular basis $X_N^{\nu}$, $X_N^{-\nu-1}$ of solutions to the homogeneous RW equation \eqref{eqn:RW} with asymptotic behavior

\begin{subequations}
	\begin{align}
	&X_N^{\nu} \to \cos{(\omega r_* + \alpha_{\nu})}, && r_* \to \infty,\\
	&X_N^{-\nu-1} \to \cos{(\omega r_* + \alpha_{-\nu-1})}, && r_* \to \infty,
	\end{align} \label{eqn:XNasymptotic}
\end{subequations}
for some constants $\alpha_{\nu}$, $\alpha_{-\nu-1}$, described explicitly in \eqref{eqn:alphanu}. Here, $\nu$ indicates the renormalized angular momentum, which reduces to $\ell$ in the low frequency limit \cite{sasaki2003}. Naively, one could therefore try to regard $X_N^{\nu}$, $X_N^{-\nu-1}$ respectively as the $\propto r^{\ell}$ tidal field and the $\propto r^{-\ell-1}$ multipolar response but it is not straightforward to connect that point of view to the actual radiation zone asymptotic behavior \eqref{eqn:XNasymptotic}. Nevertheless, it was shown in \cite{Chakrabarti2013} that, defining $a(\omega)$ through

\be
X_{\ell m \omega}^{\text{reg}} \propto X_N^{\nu} + (2 M_1 \omega)^4 a(\omega) X_N^{-\nu-1},
\label{eqn:chakrabartiasympt}
\ee
$\tilde{F}(\omega)$ can be written in a $2M_1\omega \ll 1$ expansion as
\be
\frac{3G}{4M_1^5}\tilde{F}(\omega) = -\frac{428}{7}a(\omega)-\frac{56}{107} +\cO((M_1\omega)^2).
\label{eqn:Fifa}
\ee
This is the leading order of equation (15) in \cite{Chakrabarti2013}.
Comparing \eqref{eqn:chakrabartiasympt} with our previous expressions \eqref{eqn:asymptotic}, \eqref{eqn:reflectivityBC} and \eqref{eqn:Kmark} we find that $a$ is related to $\cK$ by
\be
\begin{aligned}
	&\cK = \frac{(e^{i\alpha_{\nu}}-\cR_{\text{BH}} e^{-i\alpha_{\nu}})}{(e^{-i\alpha_{\nu}}+(2 M_1 \omega)^4 a e^{-i\alpha_{-\nu-1}})\cT_{\text{BH}}}\\
	&\qquad+\frac{(2 M_1 \omega)^4 a (e^{i\alpha_{-\nu-1}}-\cR_{\text{BH}} e^{-i\alpha_{-\nu-1}})}{(e^{-i\alpha_{\nu}}+(2 M_1 \omega)^4 a e^{-i\alpha_{-\nu-1}})\cT_{\text{BH}}}\,.
\end{aligned}
\label{eqn:Keqn}
\ee
It is important that we are assuming a hierarchy of scales in which, even though $M_1\omega \ll 1$, the orbital frequency is potentially of similar order as the resonant mode $\omega \sim \omega^{\rm QNM}_{n \ell}$. In particular, it would not be appropriate to expand $\tilde{F}(\omega)$, as in the adiabatic limit, where it could be essentially replaced by the (quadrupolar) tidal Love number \eqref{eq:linearresponse_taylor} as this would be the lowest order in the expansion with respect to the internal fundamental frequency scale $\omega/\omega^{\rm QNM}_{n \ell}$ as opposed to $M_1\omega$. 
For a black hole
\be
(2 M_1 \omega)^4 a_{\text{BH}} = -\frac{e^{i\alpha_{\nu}}-\cR_{\text{BH}} e^{-i\alpha_{\nu}}}{e^{i\alpha_{-\nu-1}}-\cR_{\text{BH}} e^{-i\alpha_{-\nu-1}}}\,,
\ee
such that we can write
\be
\cK = \frac{(2 M_1 \omega)^4 (a-a_{\text{BH}}) (e^{i\alpha_{-\nu-1}}-\cR_{\text{BH}} e^{-i\alpha_{-\nu-1}})}{(e^{-i\alpha_{\nu}}+(2 M_1 \omega)^4 a e^{-i\alpha_{-\nu-1}})\cT_{\text{BH}}}\,,
\ee
and finally, for $M_1 \omega \ll 1$, using \eqref{eqn:Fifa} and low frequency expansions for black hole quantities such as $\cR_{\text{BH}}$, $\cT_{\text{BH}}$,\footnote{Note that $a-a_{\text{BH}}$ can be replaced by $\tilde{F}-\tilde{F}_{\text{BH}}$ in a low frequency expansion according to \eqref{eqn:Fifa} and, in the same expansion, $\tilde{F}_{\text{BH}}$ could be replaced by the associated adiabatic tidal deformability since there for the black hole $\omega^{QNM}_{n \ell} M_1 \sim 1$. Now the tidal Love number of the black hole vanishes leaving a proportionality with $\tilde{F}$.} derived for instance in \cite{poisson1995}, and reviewed for convenience in Appendix \ref{app:near}
\be
\cK= C_1 (2 M_1 \omega)^2 \frac{3G \tilde{F}(\omega)}{4M_1^5}+\cO(M^3_1 \omega^3)
\label{eqn:Llowf},
\ee
%C1 = -1995/428
with $C_1$ a numerical factor. We make the derivation of \eqref{eqn:Llowf} more explicit in Appendix \ref{app:near} and explain there why we cannot determine the constant $C_1$ a priori. Essentially, it is due to a regularization dependence in \eqref{eqn:Fifa}. We instead fix the coefficient to be $C_1=-1/12$ by matching the energy flux to the post-Newtonian expansion. In conclusion, at least in the limit $M_1 \omega \ll 1$, the near-zone characterization of the object through $\cK$ is straightforwardly related to the effective field theory description using $\tilde{F}(\omega)$.
\subsection{Enhanced GW flux at resonance}
\label{sec:enhancedflux}
We will now describe how there is also an increase in gravitational wave flux at resonance. This is in particular what was studied in \cite{Cardoso:2019nis} and found there to give rise to an ``unobservable high-frequency glitch". We stress that this is not the same as looking for a phase shift like \eqref{eqn:PS}, which does not try to directly measure the variation in gravitational waves. Rather it considers the impact on the orbit and the indirect change in GW phasing associated to this.  The same authors also consider the impact of a phase shift in the gravitational wave signal, but only due to the additional energy lost through this enhanced gravitational wave flux during resonance. However, this is not the same as the total energy lost from the orbit at resonance. Even in the simple example of \cite{Cardoso:2019nis} where no energy is absorbed by the ECO itself, such that eventually all energy in the excitation must be dissipated through gravitational waves. The reason is that this dissipation does not happen instantly at resonance. In fact, the more sharply peaked the resonant frequency, the longer it will take for all this absorbed energy to be radiated. It is therefore more appropriate to consider how much energy goes into the excited mode at resonance, as was done in section \ref{sec: HO}, as opposed to how much of this is immediately radiated at resonance.  Nevertheless, the present formalism allows for a perspicuous description of the peak in gravitational wave flux such that we shall take a moment to describe it. 

In an extreme mass ratio limit $q \ll 1$, the source $T_{\ell  m}({\omega},{r})$ in \eqref{eqn:RW} can be constructed from the stress-energy tensor of an orbiting point-particle $M_2$. If this companion is on a circular orbit with angular frequency $\dot{\phi}$, the total gravitational wave luminosity is given by \footnote{This is normally expressed in terms of the analogues of \eqref{eqn:Zinfty} and \eqref{eqn:ZH} starting from the Bardeen-Press-Teukolsky equation \cite{bardeen1973, teukolsky1973} which is more convenient in this setup. Nevertheless, treating $T_{\ell m}$ carefully, one should also be able to find them from \eqref{eqn:Zinfty} and \eqref{eqn:ZH} using the Chandrasekhar transformation \cite{chandrasekhar1975} and we do so to avoid introducing unnecessary new notation. }

\be
\dot{E}^{\infty} = \sum_{\ell m} \frac{4 (m \dot{\phi})^2}{\pi}\left|Z^{\infty}_{\text{BH}}\right|^2 \left| 1  + \cK  \frac{Z^{H}_{\text{BH}}}{Z^{\infty}_{\text{BH}}}\right|^2.
\label{eqn:Eflux}
\ee

\begin{figure}
	\includegraphics[width=.45\textwidth]{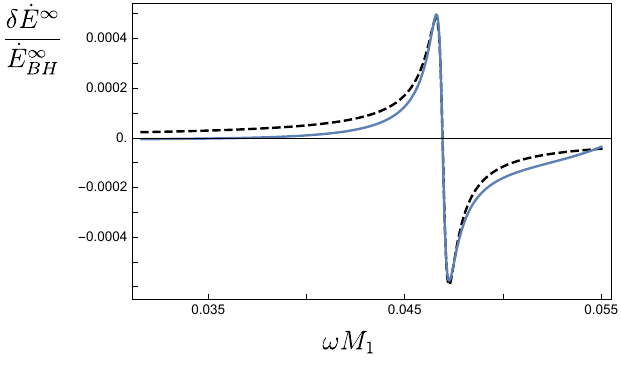} \\
	\includegraphics[width=.45\textwidth]{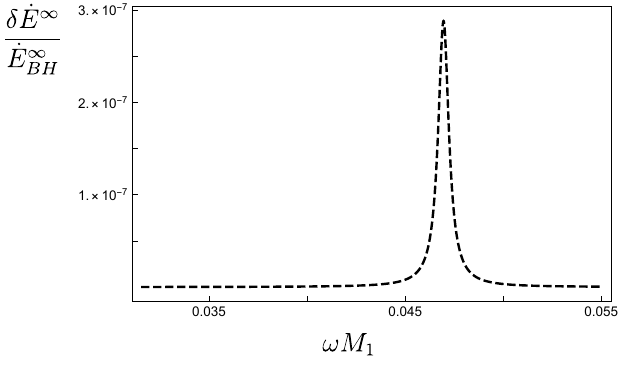} \\
	\caption{At resonant orbital frequencies there is an enhanced gravitational wave flux $\delta \dot{E}^{\infty}=\dot{E}^{\infty}-\dot{E}^{\infty}_{\rm BH}$ associated to an ECO with respect to a black hole $\dot{E}^{\infty}_{\rm BH}$, as found from \eqref{eqn:Eflux} (top, blue, full) and compared to the approximation at resonance \eqref{eqn:deltaEflux} (top, black, dashed). However, the cross term from \eqref{eqn:deltaEflux} or ``interference term'', which does not lead to a net enhancement of flux, can, as in the illustrated example, dominate strongly over the Breit-Wigner peak associated to that same resonance (bottom, black, dashed).}
	\label{fig:resonance}
\end{figure}

Consider the normalized difference of $\dot{E}^{\infty}$ with respect to the expression for a black hole around a particular QNM frequency $\omega^{\rm QNM}_{n \ell}$ for a certain $(\ell, m)$ mode
\begin{subequations}
	\be
	\begin{aligned}
		\frac{\delta \dot{E}^{\infty}}{\dot{E}^{\infty}_{BH}} &= \left|\frac{\cA}{m \dot{\phi}-\omega^{\rm QNM}_{n \ell}}\right|^2\\
		&\qquad+2\: \mathrm{Re}\left(\frac{\cA}{m \dot{\phi}-\omega^{\rm QNM}_{n \ell}}\right) ,
	\end{aligned}%
	\label{eqn:deltaEflux}
	\ee%
	with
	\be
	\cA = \left[ (\omega-\omega^{\rm QNM}_{n \ell})\cK\frac{Z^{H}_{\text{BH}}}{Z^{\infty}_{\text{BH}}} \right]_{\omega=\omega^{QNM}_{n \ell}}.%
	\ee%
\end{subequations}
The form of \eqref{eqn:deltaEflux} consists of a typical Breit-Wigner peak with an additional interference term. A representative example of this form is shown in figure \ref{fig:resonance}. The result is not surprising and it has been observed in particular cases \cite{Pons2002,Pani2010,Macedo2013}. The given example highlights that often in these resonances there seems to be a large contribution from the asymmetric interference term in \eqref{eqn:deltaEflux}.  It should be noted that although we have discussed a resonance in the radiated energy, there is also an interesting conservative resonant response in the backreaction on the orbit \cite{isoyama2016}. Moreover, once the full self-force, including the conservative piece, is known for the black hole, it is evident that one can simply add to this the force associated to the (homogeneous) last term in \eqref{eqn:Rsol}, which requires no additional regularization \cite{drivas2011dependence}. We shall not pursue this point further here.

To intuitively understand this enhanced gravitational wave flux better, let us connect this general extreme-mass ratio formula to its post-Newtonian limit. 
From the standard low frequency $\dot{\phi} M_1 \ll 1$ expansions for the black hole quantities, as can be found for instance in \cite{poisson1995}, in addition to \eqref{eqn:Fifa}, one finds that the energy flux \eqref{eqn:Eflux} can be expressed in this limit as
\be
\dot{E}^{\infty} = \sum_{\substack{\ell=2\\ m=\pm2}} \frac{4 (m \dot{\phi})^2}{\pi}\left|Z^{\infty}_{\text{BH}}\right|^2 \left|1+ (M_1\dot{\phi})^{10/3} \frac{3G \tilde{F}(m \dot{\phi})}{2M_1^5}\right|^2.
\label{eqn:EfluxHO}
\ee
Expanding this result further, assuming one is not too close to resonance as to avoid subleading terms including $\tilde{F}(m \dot{\phi})$ to become significant, one finds
\begin{equation}
\begin{aligned}
&\dot{E}^{\infty} = -\frac{32}{5} \left(\frac{M_2}{M_1}\right)^2 \left(M_1 \dot{\phi}\right)^{10/3}   \\  &\quad\times \left[ \frac{\dot{E}^{\infty}_{BH}}{\dot{E}_{N}}+3 \frac{\mathrm{Re}(\tilde{F}(2 \dot{\phi}))}{M_1^5} \left(M_1 \dot{\phi}\right)^{10/3} \right],
\end{aligned}
\label{eqn:EfluxHO2}
\end{equation}
where we have made explicit only the leading order $\tilde{F}$ contribution, despite the existence of many additional terms leading with respect this. These, denoted by $\dot{E}^{\infty}_{BH}$ where $\dot{E}^{\infty}_{BH}(\omega M_1 \to 0) \to \dot{E}_{N}$, would however, simply be the black hole result which are known to very high order, see \cite{sasaki2003}, for a review. \eqref{eqn:EfluxHO2} can be readily compared to the post-Newtonian expression of GW luminosity for a binary including effects of tidal deformability by using \eqref{eq:linearresponse_lovenumbers} in the adiabatic limit
\be
\begin{aligned}
	&\dot{E}^{\infty} = -\frac{32}{5} \left(\frac{M_2}{M_1}\right)^2 \left(M_1 \dot{\phi}\right)^{10/3} \\
	&\quad\times\left[ \frac{\dot{E}^{\infty}_{BH}}{\dot{E}_{N}}+3 \frac{\mu_2}{M_1^5} \left(M_1 \dot{\phi}\right)^{10/3} \right].
\end{aligned}
\label{eqn:EfluxHO3}
\ee
For instance, with $\nu = \frac{M_1 M_2}{(M_1+M_2)^2}$ and $M=M_1+M_2$ \cite{flanagan2008} 
\begin{align}
&\dot{E} = -\frac{32}{5} \nu^2 \left(M \dot{\phi}\right)^{10/3} \label{eq:PNPower}\\ &\times \left[  \frac{\dot{E}^{\infty}_{BH}}{\dot{E}_{N}}+3\frac{M_1+3M_2}{M_1} \frac{\mu_2}{M^5} \left(M \dot{\phi}\right)^{10/3} + 1 \leftrightarrow 2 \right]. \nn
\end{align}

Aside from the fact that \eqref{eqn:EfluxHO2} is valid only as $q\ll 1$, the difference is that \eqref{eqn:EfluxHO2} still captures some of the dynamic aspects of the tidal deformability. It suggests a way of resumming dynamic tidal interactions in a post-Newtonian expansion and resonates well with the introduction of an effective tidal Love number in \cite{steinhoff2016}. In that paper, $\mu_2$ was replaced by an effective-Love number function in order to efficiently capture dynamic tidal effects in an effective-one-body approach
\be
\mu_2 \to \mu_{\text{eff}}=- \frac{E_{ab}Q^{ab}}{E_{cd}E^{cd}},
\ee
similar to the response function in \eqref{eq: quadrupole linear response}. It would be interesting to investigate how this works out precisely but we shall not pursue this further here.  \\
We can again observe from \eqref{eqn:EfluxHO2} that the enhanced gravitational wave flux at resonance, as expected, is simply because the deformed object has a varying quadrupole moment itself which also contributes to the emitted gravitational waves. This result was derived for a fixed circular orbit. It does not allow us to conclude that, with an evolving orbit, this deformation would simply disappear as the resonance is past. Therefore, it would be incorrect to conclude that all the energy lost to the orbit was simply the additional flux during resonance. Instead, off-resonance, the excited mode will continue to ring down. In itself, this will be too weak to observe directly but in the total energy that was lost to the orbit when the associated energy was transferred to the mode, a measurable phase shift might have been induced into the waveform. This is what we will investigate now.

\section{Detectability} \label{sec: dectectability}

To at least get an order of magnitude estimate on the detectability, we will perform a Fisher analysis pinned down to a particular ECO model. The ECO model is the simplest reflecting shell model \cite{Cardoso:2019rvt} which serves our illustrative purposes although a similar analysis on different models might also be performed.

\subsection{The model}

The reflecting shell model  is decisively simple, but it nevertheless captures the physics of horizon absence. Phenomenologically, this absence inevitably introduces a certain amount of reflection of incoming waves within the light ring. The simplest model that captures this, is a purely reflecting surface put at a Schwarzschild coordinate $r_{0}=2M_1(1+\epsilon)$ where $M_1$ is the mass of the object and $\epsilon$ is a small dimensionless parameter that quantifies the ``closeness'' to a black hole spacetime as introduced in section \ref{sec: Introduction and summary}. The proper distance from the Schwarzschild radius $r=2M_1$ to the reflective surface at $r_{0}$ in the small $\epsilon$ limit is given by
\begin{equation}
\delta = 4M_1\epsilon^{1/2} + \mathcal{O}(\epsilon^{3/2})\,.
\end{equation}
The parameter $\epsilon$ ranges from $\epsilon_{\rm min}\approx10^{-80}$ where $\delta$ is about one Planck length to $\epsilon_{\rm max}\approx0.0165$ where the object has a clean photonsphere \cite{Cardoso:2017njb}. However, in practice our upper limit will be $\epsilon_{\rm max}\approx10^{-5}$. This will be explained below.

On the level of the wave equation, the surface can be treated effectively by using reflective boundary conditions at $r_{0}$ and the QNMs in the $\epsilon\to0$ limit, valid in our $\epsilon$ range, have been analyzed by Cardoso et al. \cite{Cardoso:2019nis}
\be
\begin{aligned}
	M_1&\omega_{n\ell}^{\rm QNM} = \frac{n\pi}{2|\log\epsilon|} -\frac{i}{|\log\epsilon|^{2\ell+3}} \\ &\times \left[\frac{(2n\pi)^{\ell+1}\Gamma(\ell+1)\Gamma(\ell-1)\Gamma(l+3)}{4\Gamma(2\ell+1)\Gamma(2\ell+2)}\right]^2 . \label{eq: QNM frequency} 
\end{aligned}
\ee
Furthermore, it is also shown that the Love numbers of those objects in the $\epsilon\to0$ limit take the following approximate form \cite{Cardoso2017(2)}
\begin{equation}
k_\ell \approx \frac{1}{a_\ell + b_\ell\log\epsilon},\label{eq.:lovenumberlog}
\end{equation}
where $a_{\ell},b_\ell$ depend on the multipolar index $\ell$. Note that $\epsilon \ll 1$, the Love numbers are negative. For instance, for quadrupolar deformations for our shell model, it has been found that $a_2=35/8$ and $b_2=15/8$ \cite{Cardoso2017(2)}.  Using this approximation, the phase shift \eqref{eq:DeltaPhi_stars2(2)} is fully determined by the location of the surface $\epsilon$, the mass ratio $q$ and the multipolar indices $(\ell,m)$. In particular, from \eqref{eq: QNM frequency}, \eqref{eq.:lovenumberlog} and \eqref{eq:DeltaPhi_stars2(2)}, the mass dependence scales to leading order in $|\log\epsilon|$ as
\begin{equation}
\Delta\Phi\propto q^{-1}(1+q)^{-\frac{1}{3}(2\ell-1)}|\log\epsilon|^{-\frac{4}{3}\ell+\frac{5}{3}}, \label{eq: scaling PS}
\end{equation}
where the proportionality constant, which only depends on $(\ell,m)$, is about $0.4$  for quadrupolar waves and increases slightly with $\ell$. From \eqref{eq: scaling PS}, it immediately follows that for small mass ratios, the phase shift is inversely proportional to the mass ratio  and the $\epsilon$-dependence is such that higher order $\ell$-modes are at least suppressed by factors of $|\log\epsilon|^{\frac{4}{3}}$. In particular for our upper limit $\epsilon_{\rm max}\approx 10^{-5}$, it can be computed that higher order $\ell$-modes are generically suppressed by at least two orders of magnitude
\begin{equation}
\frac{\Delta\Phi(\ell+1)}{\Delta\Phi(\ell)}\lesssim 10^{-2}. \label{eq: higher order PS}
\end{equation}
This makes that the largest phase shift comes from quadrupolar excitations, by \eqref{eq: scaling PS} it can be seen that the phase shift itself has a very weak $\epsilon$-dependence. In particular, when $\epsilon$ runs from Planck scale ($\epsilon\sim10^{-80}$) to roughly $\epsilon\sim 10^{-5}$, the phase shift only changes by two orders of magnitude.

\subsection{Criteria for detectability} \label{subsec:criteria}
In order to detect the phase shift, the following criteria should be fulfilled:
\begin{enumerate}[label=(\alph*)]
	\item Resonant excitation should take place at a frequency lower than that at the moment of merger,
	\item The resonant frequency should lie within the detector band,
	\item The measurement error on $\Delta\Phi$ should be smaller than $\Delta\Phi$ itself.
	
\end{enumerate}

\begin{figure*}[ht!]
	\begin{center}
		
		\includegraphics[width=0.6\textwidth]{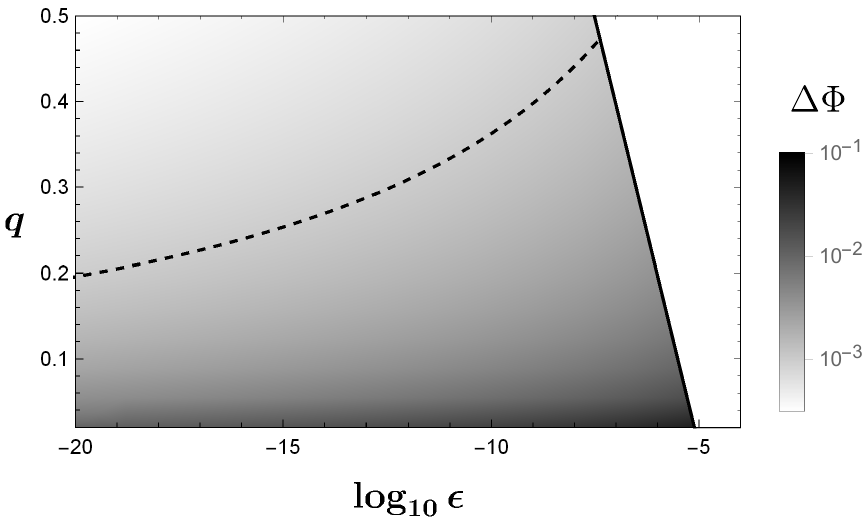}
	\end{center}
	\caption{Value of the phase shift $\Delta \Phi$, given by (\ref{eq:DeltaPhi_stars2}), as a function of the mass ratio $q$ and the dimensionless parameter $\epsilon$, which measures how close the ECO surface is to the would-be horizon via $r_{0}=2M_1(1+\epsilon)$. The solid line corresponds to the merger of the binary system, as indicated by \eqref{eq:mergerfrequencycondition}. The resonance takes place before merger for points to the left of this line and therefore satisfy criterium (a) of \ref{subsec:criteria} for detectability. The dashed line corresponds to $\Delta \Phi= 10^{-3}$ rad; all points below this line correspond to a phase shift greater than this value.}
	\label{DeltaPhiPlot}
\end{figure*}

The first condition is purely the statement of the resonant effect happens at all, the others are detector dependent statements. The first condition puts a restriction on $\epsilon$ in terms of the mass ratio. In particular if one estimates that the plunge initiates at an orbital separation of roughly $r=6(M_1+M_2)$, then the requirement that the orbital resonant frequency, associated to \eqref{eq: QNM frequency}, is smaller than the approximate merging frequency, $\omega_{\rm max} \approx (M_1+M_2)^{-1} 6^{-3/2} $, translates to
\begin{equation}
q<\frac{4}{\pi 6^{3/2}}|\log\epsilon|-1\,. 
\label{eq:mergerfrequencycondition}
\end{equation}
This also puts the constraint that $\epsilon\lesssim  10^{-5}$ corresponding to $q=0$. The second condition requires $\epsilon$ to be constrained by
\begin{equation}
\frac{1}{4 M_1 f_\text{max}} < |\log \epsilon | < \frac{1}{4 M_1 f_\text{min}}\,,
\end{equation}
or reinstating factors of $G_N, c$ and plugging in the detector band $(f_\text{min}=1 \, \text{Hz}, f_\text{max}=10^4\, \text{Hz})$ of the Einstein Telescope:
\begin{equation}
10^{-1}\:\left(\frac{50 M_{\odot}}{M_1}\right)\lesssim |\log\epsilon| \lesssim 10^3\:\left(\frac{50 M_{\odot}}{M_1}\right)\,.\label{eq:}
\end{equation}
Using the constraint set by \eqref{eq:mergerfrequencycondition}, we find that the central mass should fit within $0.4 \:M_\odot \lesssim M_1 \lesssim 4000M_\odot$. Conversely, this means that the full range of current ECO models, ranging from $|\log \epsilon| \sim 1$  with structure near the light-ring to Planck-scale from horizon for which $|\log \epsilon| \gtrsim 2\times 10^2$ can give resonances within band for mass ranges accessible with current ground-based observatories. Note that to reach such microscopic scales requires $M_1$ to be smaller than $\sim 250 M_\odot$. The conclusions on the conditions (a)-(b) are summarised in Figure \ref{DeltaPhiPlot}, which depicts the value of the phase shift $\Delta \Phi$ as a function of mass ratio $q$ and ECO radius $r_0(\epsilon)$, as well as the line at which merger happens.

Finally, the bounds set so far on the mass $M_1$ and on $\epsilon$ can be used to check under what conditions the third requirement (c) is met.  The methodology used is that of the Fisher matrix formalism, which allows the calculation of the variance $\sigma(\theta)$ with which the value of a parameter $\theta_i$ can be determined from a given signal, given the sensitivity curve $S_h(f)$ of a detector and the Fourier transform $\tilde{h}(f,\theta_1, \dots, \theta_n)$ of the theoretical prediction of the gravitational wave. This formalism is extensively used to make predictions for how accurate (future) detectors will be able to measure observables $\theta_i, \dots, \theta_n$,\cite{Maggiore}. \newline
The formalism states that the Fisher matrix $\Gamma_{ij}$, defined as
\begin{equation}
\Gamma_{ij} = 4\, \text{Re} \int \left(\frac{\partial \tilde{h}(f)}{\partial \theta_i} \right)\, \left(\frac{\partial\tilde{h}(f)}{\partial \theta_j}\right)^* \frac{df}{S(f)} \,,
\end{equation}
where $S(f)$ is the power spectral density of the detector under consideration, leads to the covariance matrix $\Sigma_{ij}$,
\begin{equation}
\Sigma_{ij} =\Gamma^{-1}_{ij}\,,
\end{equation}
in which each element is the covariance of two observables $\theta_i, \theta_j$. In particular, the square root of the diagonal elements of the covariance matrix are the standard deviations $\sigma(\theta_i)$ of each of the observables $\theta_i$, and are therefore a direct measure of the accuracy with which the detector will be able to determine their values. 

In what follows, we apply the Fisher matrix formalism to calculate the relative errors $\frac{\sigma(\Delta \Phi)}{\Delta \Phi}$, \emph{i.e.} the ratio of the standard deviation of the phase shift and the phase shift itself, for a range of ECO-models and a range of mass-ratios of the binary system. The gravitational waveform $\tilde{h}(f,\theta_1, \dots, \theta_n)$ used is given by 
\begin{equation}
\tilde{h}= A\, f^{-7/6}\, e^{i \Phi}\,,
\label{TaylorF2}
\end{equation}
in which the phase $\Phi$ is given by (\ref{FlanaganEquation}), the phase shift $\Delta\Phi$ due to the resonance is given by (\ref{eq:DeltaPhi_stars2}), and we have taken $\Phi_{\rm pp}$ to be the TaylorF2 approximant to 2.5 PN order, as given in \cite{approximant}. This approximant allows for the two component masses to have spin (that are taken to be aligned with the orbital angular momentum), but otherwise models them as featureless point particles. The resulting spin-orbit coupling and spin-spin coupling, respectively, are taken into account to 2.5 PN and 2 PN order. This approximant excludes non-resonant tidal forces. Those effects have already been investigated in \cite{Cardoso:2017cfl} and excluding them is not expected to change the order of magnitude estimations provided by the Fisher matrix formalism. In the Fisher calculations to follow, the relative error is calculated for the coalescence time $t_c$, coalescence phase $\phi_c$, chirp mass, dimensionless mass ratio, and the phase shift. The value for the amplitude $A$ in (\ref{TaylorF2}) is fixed by the choice of SNR. The upper cut-off frequency of the Fisher matrix calculation is taken to be the frequency corresponding to the ISCO of $r=6(M_1+M_2)$, whereas the lower cut-off frequency is dependent on the gravitational wave detector under consideration.  \newline
The calculation is done for Einstein Telescope using the power spectral density $S(f)$ as given in \cite{PSD_ET}. A similar Fisher calculation was performed using the power spectral density of the Advanced LIGO network, but the results showed that the relative error of the phase shift was much larger than unity for all phase space considered, leading to the conclusion that the current generation detectors are not able to measure the ECO shift. This result is corroborated by a recent paper \cite{NikhefResonances} in which the binary black hole systems of GWTC-1 were analysed for resonances, and none were found. The results below will therefore focus on the Einstein Telescope. As such, the lower cut-off frequency in the definition of the Fisher matrix is taken to be $2 \text{ Hz}$.

\subsection{Results}

\begin{figure}
	\centering
	\includegraphics[width=.40\textwidth]{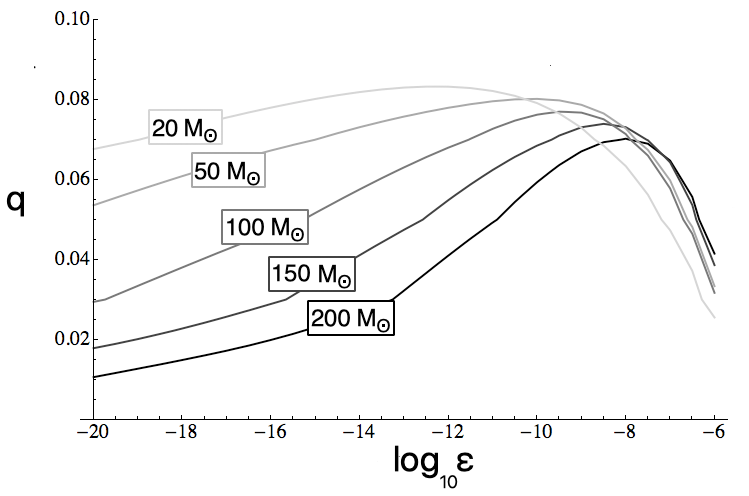} \\
	\includegraphics[width=.50\textwidth, clip=true, trim=2cm 0cm 3cm 0cm]{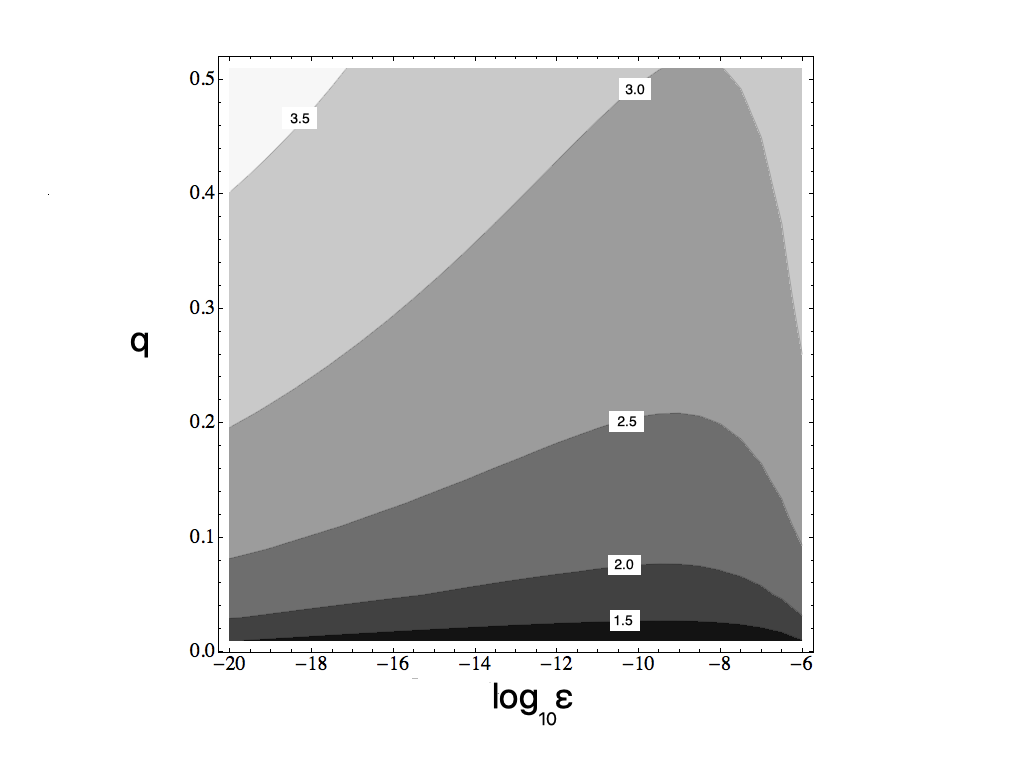} \\
	\caption{Contours of relative error $\sigma(\Delta \Phi)/\Delta \Phi$ of the phase shift, induced by a resonance in an exotic compact object, as measured by ET, calculated by the Fisher matrix formalism, over a range of the ECO radius $r_{0}=2M_1(1+\epsilon)$ and for mass ratio $q$ of the binary system at SNR $ = 400$. The relative error is inversely proportional with SNR such that the contour values scale accordingly. \newline \textit{Top}: $\sigma(\Delta \Phi)/\Delta \Phi = 100$ for different values of the mass $M_1 = 20,50,100,150,200\, M_\osun$ of the ECO, with lower relative errors below the contour. \newline \textit{Bottom}: $\log_{10}\left(\sigma(\Delta \Phi)/\Delta \Phi \right)$ as indicated on the contours, for an ECO of mass $M_1 = 100\,M_\osun$. }
	\label{fig:Detectabilities}
\end{figure}

Figure \ref{fig:Detectabilities} shows the results for the relative errors on the phase shift for binary systems in which only the heavier of the two compact objects undergoes the resonance and is given a fixed mass $M_1$, while the companion compact object's mass $M_2$ is varied via the mass ratio $q = M_2/M_1$ of the binary system. This is done for five different fixed values for the heavier compact object, $M_1 = 20,\, 50,\,100,\, 150,\,  200 M_\odot$. \newline
We take the regions within the contour line of unit relative error to be measurable by the detector. For the different ECO masses ($M_1$) considered, we find that, for the phase space considered, the resonance takes place well before the merger. 
However, these results show that for the most likely events for the Einstein Telescope (SNR $\sim$ 400), the resonances will be undetectable. Nevertheless, for unique events (SNR $\sim$ 4000, or a luminosity distance of order $\sim 10$Mpc), the ECO resonance shifts for binaries of sufficiently asymmetric masses might be detected. In that case, we find that the relative error of the phase shift $\Delta \Phi$ is smaller than unity over many orders of magnitude of $\epsilon$. This wide range of $\epsilon$ is expected based on the weak (logarithmic) dependence of the phase shift on this parameter. We find that the width of this range increases for decreasing value of the ECO mass, and for increasingly asymmetric mass ratios. We find small relative errors exactly in the parameter range for which the phase shift \eqref{eq: scaling PS} is large, namely for small mass ratios $q$ and relatively large values of $\epsilon$. Moreover, it reaches a minimal value for a  binary of a given mass ratio. This minimal relative error is reached at larger $\epsilon$ for increasing ECO mass.
Given these results, most notably the required small value for the mass ratio $q$ and the fact that the companion star is expected to have a mass $M_2 > 1.4\ M_\odot$, candidates of ECOs that could be detected by the Einstein Telescope should have a mass $M_1$ of several tens to a hundred solar masses as well as be close enough to supply the required high SNR. 
\newline
We have tested the robustness of these results by changing the values of the dimensionless spins of the binary system and have found the order of magnitudes of the relative error to be largely independent of such changes. We have also tested the robustness to variations in the cut-off frequency, to account for the fact that the ISCO-frequency of rotating black holes is not at $6^{-3/2}(M_1+M_2)^{-1}$, as well as this only being a test-particle approximation. Over the expected range of such corrections to the merging frequency, described for instance in \cite{favata2011conservative}, we find that the relative errors change by at most a factor of order $\sim 3$, with relative errors decreasing for higher cut-offs and increasing for lower cut-offs. We therefore conclude that the results presented are representative.

\section{Outlook}
\label{sec: outlook}

We derived an expression for the phase shift due to a tidal resonance in the Newtonian limit for circular orbits and spinless components and tested it in a Fisher analysis for the specific model of a reflecting surface. We believe that the dependence on the distance from the horizon and mass ranges for which the phase shift is significant are a robust order of magnitude estimate, that is more generally applicable:    

First, we have also performed Fisher matrix calculations including aligned component spins as well as varying cut-off frequencies and the outcomes agree to the same order of magnitude. We believe this gives a qualitative indication that a similar phase shift can play a similar role also when other complications such as eccentric orbits are taken into account. Nevertheless, based on our results, it would be interesting to perform a more detailed analysis for smaller mass ratio systems. In particular, it would be interesting to perform a study of the effects for EMRI systems with LISA-targeted wave-forms in the future.

Second, the toy model of purely reflecting boundary conditions near the horizon allows to have a particular value of the phase shift depending on a minimum of parameters, but the method is more generally valid. For other objects for which the Love numbers and QNMs scale with redshift $z \sim \log \epsilon$ in a similar way as for the surface $k \sim 1/\log \epsilon$ and $M \omega \approx \log \epsilon$, we expect a similar outcome.
In general however, the reflection coefficient $\mathcal R$ of a compact object will be smaller than unity, and therefore the phase shift we derived is likely an upper bound on the effect caused by resonant excitation of the fundamental QNM in those models. It would be interesting to investigate more general models of compact objects and run a Fisher analysis including the QNMs and Love numbers as free parameters, instead of keeping them fixed by the relation to $\log \epsilon$. 

Third, we focused on resonant excitation of the fundamental quadrupolar frequency $\omega_{n \ell m} = \omega_{022}$. Other modes can play a role as well but the value of the phase shift is maximal for the quadrupole modes $\ell = 2$. Higher multipoles $\ell$ give a smaller contribution to the phase shift, as was indicated in \eqref{eq: higher order PS}. 

The study of overtone numbers through our method of replacing the overlap integral would involve going beyond the low frequency limit, and include besides the Love numbers also other tidal constants appearing in the low-frequency expansion of the linear response. To do this confidently, it should be quantified how strongly the low-frequency expansion coefficients are dominated by the lowest lying modes. In first instance, this means scrutinizing the assumption we have made that the contribution of higher order modes to the Love number is negligible, mainly on the argument that this should give the correct order of magnitude. We leave that to future work.

%\section*{Acknowledgments}
\begin{acknowledgments}
 We thank Archisman Ghosh, Thomas Hertog, Tanja Hinderer, Anuradha Samajdar, Maarten van de Meent, Chris Van den Broeck for discussions, the organizers of the Dutch multimessenger astronomy in the 2030s for the stimulating atmosphere that started part of this work. R.T, K.F. and B.V. thank the University of Maastricht for hospitality. R.T.\ thanks Nikhef for hospitality during the initial stages of this work. This work makes use of the Black Hole Perturbation Toolkit. K.F.\ is Aspirant FWO-Vlaanderen (ZKD4846-ASP/18). B.V.\ was partially supported by the National Science  Foundation  of Belgium (FWO) grant G.001.12, the European Research Council grant no.\ ERC-2013-CoG 616732 HoloQosmos, the KU Leuven C1 grant ZKD1118 C16/16/005, the FWO Odysseus grant G0H9318N and the COST action CA16104. The  work  of  RT is supported in part by by the European Research Council grant no. ERC 616732 -- HoloQosmos, bijzonder onderzoeksfonds C16/16/005 -- horizons in hoge-energie fysica and COST Action CA16104 \emph{GWVerse}.
 
 This paper is dedicated to the memory of Frederik Goelen,
former PhD student at KU Leuven and contributor to the early stages of this work.
\end{acknowledgments}

\appendix
\numberwithin{equation}{section}
\begin{widetext}
\newpage
\section{Low frequency expansions for BH perturbations}
\label{app:near}

In Section \ref{sec:nearzone}, we have given the expression \eqref{eqn:Llowf} which relates, in the low frequency limit $m_1 \omega \ll 1$, a near-zone characterization of an ECO in terms of $\cK$ to the EFT characterization in terms of a tidal response function $\tilde{F}(\omega)$. In this appendix we will elucidate the derivation of this result. Our starting point is \eqref{eqn:Keqn}. To make this more explicit we first of all use the appropriate expansions for $B_{\text{in}}$ and $B_{\text{out}}$ from which one can derive, $\cR_{\text{BH}}$ and $\cT_{\text{BH}}$. We find these expressions combining 

\be
A_{\text{in}} = \frac{(2\ell)!(2\ell+1)!!}{2(\ell-2)!(\ell+2)!}(\frac{i}{2m_1 \omega})^{\ell+1}e^{-i2m_1 \omega (\log{4m_1\omega}-\tau_{\ell}-\beta_{\ell})}(1-\pi m_1 \omega + O((m_1 \omega)^2))\,,
\ee

\be
A_{\text{out}}= \frac{(2\ell)!(2\ell+1)!!}{2(\ell-2)!(\ell+2)!}(\frac{-i}{2m_1 \omega})^{\ell+1}e^{-i2m_1 \omega (-\log{4m_1\omega}-\tau_{\ell}+\beta_{\ell})}(1-\pi m_1 \omega + O((m_1 \omega)^2))\,,
\ee

where

\be
\beta_{\ell} = \frac{1}{2}(\psi_0(\ell+1)+\psi_0(\ell)+\frac{(\ell-1)(\ell+3)}{\ell(\ell+1)})\,,
\ee

\be
\tau_{\ell} = 2\gamma + \psi_0(\ell-1)+\psi_0(\ell+3)-1\,,
\ee

from \cite{poisson1995} with 
\be
B_{\text{in}} = - \bar{A}_{\text{out}}\,,
\ee

\be
B_{\text{out}} = A_{\text{in}}\,.
\ee

For the reflection and transmission coefficients one then finds

\be
\cT_{\text{BH}} = \frac{2(\ell-2)!(\ell+2)!}{(2\ell)!(2\ell+1)!!}(\frac{i}{2m_1 \omega})^{-\ell-1}e^{i2m_1 \omega (\log{4m_1\omega}-\tau_{\ell}-\beta_{\ell})}(1+\pi m_1 \omega + O((m_1 \omega)^2))\,,
\label{eqn:TBH}
\ee

\be
\cR_{\text{BH}} =-e^{-i4m_1 \omega \tau_{\ell}}+ O((m_1 \omega)^2)\,.
\label{eqn:RBH}
\ee

To find $\alpha_{\nu}$, we use from \cite{Chakrabarti2013}

\be
\alpha_{\nu}=\frac{1}{2 i} \ln \frac{A^{\nu}_{C,\text{out}}}{A^{\nu}_{C,\text{in}}}\,,
\label{eqn:alphanu}
\ee

with

\be
A^{\nu}_{C,\text{in}}= \frac{1}{2}i^{-\nu+i2m_1\omega -1} \sum_{-\infty}^{\infty} -i^n (4m_1\omega)^{-i2m_1\omega}e^{i\pi(\nu+\frac{n}{2})}a_n^{\nu} \frac{\Gamma(n-2im_1\omega+\nu-1)}{\Gamma(n+2im_1\omega+\nu+3)}\,,
\ee

\be
A^{\nu}_{C,\text{out}}=\frac{1}{2}i^{-\nu+i2m_1\omega -1} \sum_{-\infty}^{\infty}  (4m_1\omega)^{-i2m_1\omega}a_n^{\nu} \frac{\Gamma(n-2im_1\omega+\nu-1)}{\Gamma(n+2im_1\omega+\nu+3)} \frac{\Gamma(n-2im_1\omega+\nu+1)}{\Gamma(n+2im_1\omega+\nu+1)}\,.
\ee

Here, the coefficients $a_n^{\nu}$ satisfy the recurrence relation

\be
\alpha_n^{\nu}a_{n+1}^{\nu} + \beta_n^{\nu}a_n^{\nu}+\gamma_n^{\nu}a_{n-1}^{\nu}=0\,,
\ee

with

\bea
\alpha_n^{\nu} &=& -\frac{i 2 m_1 \omega(\nu+n-2im_1\omega-1)(\nu+n-2im_1 \omega+1)(\nu+n+2im_1 \omega-1)}{(\nu+n+1)(2(\nu+n)+3)}\,, \\
\beta_n^{\nu} &=&-\ell(\ell+1)+(\nu+n)(\nu+n+1)+2(2 m_1 \omega)^2+\frac{((2m_1\omega)^2+4)(2m_1\omega)^2}{(\nu+n)(\nu+n+1)}\,, \\
\gamma_n^{\nu} &=&\frac{i 2 m_1 \omega(\nu+n-2im_1\omega+2)(\nu+n+2im_1 \omega)(\nu+n+2im_1 \omega+2)}{(\nu+n)(2(\nu+n)-1)}\,.
\eea

As is also stressed \cite{Chakrabarti2013}, the limit becomes subtle due to poles in the $\Gamma$-functions. One finds a different limit keeping $l$ generic and only setting $l \to 2$ after the expansion compared to starting with $l=2$. To get consistent results, we follow the approach from  \cite{Chakrabarti2013} which is to only let $l \to 2$ after the expansion. It should be noted however that, in any case, many of the intermediate results in \cite{Chakrabarti2013}, including \eqref{eqn:Fifa}, are regularization dependent. Therefore, it is not entirely surprising that we will find a mismatch in coefficients with the post-Newtonian result by naively combining these results. It would be interesting to redo the analysis more consistently and explore how dynamic tidal effects can be resummed in a post-Newtonian expansion but, as this is not our main goal, we postpone it to future work. \\

The renormalized angular momentum $\nu$ ensures that the minimal solutions $a_n^{\nu}$ for $n\to \infty$ and $n\to - \infty$ are compatible and is given in the low frequency expansion by \cite{Chakrabarti2013}

\be
\nu = \ell + (-\frac{(\ell-2)^2(\ell+2)^2}{2\ell(2\ell-1)(2\ell+1)}-\frac{4}{\ell(\ell+1)}+\frac{(\ell-1)^2(\ell+3)^2}{(2\ell+1)(2\ell+2)(2\ell+3)}-2)\frac{(2\omega m_1)^2}{2\ell+1} +O((\omega m_1)^4)\,.
\ee

Now from

\bea
a_0^{\nu} &=& 1\,, \nn \\
a_{1}^{\nu} &=& -\frac{2i\omega m_1 (3+\ell)^2}{2(1+\ell)(1+2\ell)} + O((m_1 \omega)^2)\,, \nn \\
a_{-1}^{\nu} &=& -\frac{2i\omega m_1 (-2+\ell)^2}{2\ell(1+2\ell)} + O((m_1 \omega)^2)\,,
\eea

one finds

%\alpha_{\nu} =\alpha_{-\nu-1}+\frac{a_{\text{reg}}}
\be
\alpha_{\nu} = (-1 - \ell) \frac{\pi}{2} + 2\omega m_1 \log{(4\omega m_1)} -2 m_1 \omega  \beta_{\ell}-\frac{20 m_1 \omega}{3(\ell-2)}+\frac{m_1 \omega (17+10\ell-3\ell^2)}{3 (1+\ell)(1+2\ell)} + O((m_1 \omega)^2)\,.
\ee

We can continue, as \cite{Chakrabarti2013}, by simply dropping the divergent term for $\ell=2$ as a type of 'minimal subtraction'. However, even if we expect this might capture the right functional form, it will not give the correct prefactor. In the main text we therefore leave this constant arbitrary, to be fixed by comparison with a post-Newtonian expansion. Nevertheless, continuing with the 'minimal subtraction', for $\ell=2$

\be
(e^{i\alpha_{-\nu-1}}-\cR_{\text{BH}} e^{-i\alpha_{-\nu-1}}) =   -4m_1 \omega \tau_{\ell} + O((m_1 \omega)^2)\,,
\ee

and we finally find \eqref{eqn:Llowf}

\be
\cK \sim - \frac{385}{1284} (2 m_1 \omega)^2 \frac{3 G \tilde{F}(\omega)}{4 m_1^5}\,.
\ee

\section{From overlap integrals to Love numbers} \label{app:effectivetheory}

We discuss the relation of the tidal constants to the overlap integrals in the Newtonian limit, adapted from \cite{Chakrabarti2013_1306}. 

The multipoles of the mass distribution, now called $\hat{Q}_{K_\ell}$ in the STF basis, are no longer independent variables but are given in terms of the mode amplitudes and overlap integrals as (cf. equation (4.9) in \cite{Chakrabarti2013_1306})
\begin{equation}
	\hat{Q}^{K_\ell} = \sum_n I_{n\ell} \hat c_{n K_\ell}.\label{eq:multipole_variables_overlapintegrals}
\end{equation}
The effective action constructed in \cite{Chakrabarti2013_1306} leads to the linear response
\begin{equation}
	\tilde Q^{K_\ell}   = - \frac 1{\ell !} \tilde{F}_\ell(\omega) {\cal F} (\hat \partial_{K_\ell} \Phi)(\omega),
\end{equation}
where $\cal F$ denotes Fourier transform. Equating those last two expressions for $\hat{Q}^{K_\ell}$ and using the solution to the harmonic oscillator equation \eqref{eq: HO} $\hat c_{n K_\ell} =  f_{n K_\ell}/(\omega_{n\ell}^2 +2i\gamma_{n\ell}\omega- \omega^2)$ and \eqref{eq: driving}, gives the linear response function:
\begin{equation}
	\tilde F_\ell(\omega) = \sum_n \frac{I_{n \ell}^2} {\omega_{n\ell}^2 +2i\gamma_{n\ell}\omega- \omega^2},\label{eq:linearresponse_multipole}
\end{equation}
Note that for the quadrupole $\ell = 2$ such a pole expansion also holds beyond Newtonian limit \cite{Chakrabarti2013_1306,Chakrabarti2013}.

The tidal deformability parameters are defined by the low frequency expansion of the response function
\begin{equation}
	\tilde{F}_\ell(\omega) = \lambda_0 + \lambda_1 \omega + \lambda_2 \omega^2 + \mathcal{O}(\omega^3),\label{eq:linearresponse_taylor}
\end{equation}
where $\lambda_0$ is the tidal deformability related to the (electric) dimensionless tidal Love number $k_\ell$, $\lambda_1$ relates to the damping, $\lambda_2$ to the response beyond the adiabatic approximation, etc. Note that often the linear response is restricted to the first coefficient in the above Taylor expansion of the linear response function. However, this is only valid far away from resonance, when $\omega \ll \omega_{n\ell}$.

Equating the two expressions \eqref{eq:linearresponse_multipole} as a low frequency expansion and \eqref{eq:linearresponse_taylor} formally allows to extract the overlap integrals as functions of the set of Taylor coefficients
\begin{equation}
	\lambda_k = \sum_n\frac{I_{n\ell}^2}{2\omega_{n\ell}^{2(k+1)}\sqrt{\omega_{n\ell}^2-\gamma_{n\ell}^2}}\left[\left(\sqrt{\omega_{n\ell}^2-\gamma_{n\ell}^2}-i\gamma_{n\ell}\right)^{k+1}-(-1)^{k}\left(\sqrt{\omega_{n\ell}^2-\gamma_{n\ell}^2}+i\gamma_{n\ell}\right)^{k+1}\right], \label{eq: lambda's}
\end{equation}
in particular, for the lowest Taylor coefficients, we have
\begin{align}
	\lambda_0=\sum_n\frac{I_{n\ell}^2}{\omega_{n\ell}^2}, && \lambda_1 = -\sum_n\frac{2i I_{n\ell}^2\gamma_{n\ell}}{\omega_{n\ell}^4}, && \lambda_2 = \sum_n\frac{I_{n\ell}^2}{\omega_{n\ell}^6}\left(\omega_{n\ell}^2-4\gamma_{n\ell}^2\right).
\end{align}
The parameter of our interest is the tidal deformability $\lambda_0$ which relates to the dimensionless tidal Love number\footnote{The Love numbers are normalized with respect to the mass of the object as in  \cite{Cardoso:2017cfl}. This makes it applicable for objects without a well-defined radius as for instance a boson star.} $k_\ell$ as
\begin{equation}
	\lambda_0 = \frac{2\ell!}{(2\ell-1)!!}k_\ell M_1^{2\ell+1}\,.
\end{equation}
As mentioned before, we will restrict to fundamental $n=1$ modes on the assumption that the associated overlap integrals are dominant above the ones with $n>1$. This implies that the sum in \eqref{eq: lambda's} can be truncated to $n=1$ only. Given this assumption, the overlap integrals can be expressed in terms of the Love numbers
\begin{equation}
	I_{1\ell} = \sqrt{\frac{2\ell!}{(2\ell-1)!!}}\omega_\ell M_1^{\ell+\frac{1}{2}}k_\ell. \label{eqapp: overlap - Love number}
\end{equation}
and response function to lowest order in $\omega$ becomes $\tilde F(\omega) = \frac{I_{1\ell}^2}{\omega_{1\ell}^2} + \mathcal{O}(\omega) = \frac{2\ell!}{(2\ell-1)!!}k_\ell M_1^{2\ell+1} + \mathcal{O}(\omega)$.

Using the result \eqref{eqapp: overlap - Love number}, allows one to rewrite the phase shift \eqref{eq:DeltaPhi_stars} in terms of the Love numbers, quasi-normal frequency and masses involved
\begin{equation}
	\Delta \Phi = \frac{25\pi N_{\ell m}^2}{6144|m|^{\frac{1}{3}(4\ell-11)}}(M_1\omega_\ell)^{\frac{4}{3}(\ell-2)}\frac{|k_\ell|}{q(1+q)^{\frac{1}{3}(2\ell-1)}},
\end{equation}
where $N_{\ell m}$ relates to $\mathcal{N}_{\ell m}$ by $N_{\ell m}^2 = 2\ell!\:\mathcal{N}_{\ell m}^2/(2\ell-1)!!$.

\end{widetext}

\bibliographystyle{toine}

\providecommand{\href}[2]{#2}\begingroup\raggedright\endgroup

\end{document}